\documentclass[aps,pre,showpacs,twocolumn]{revtex4}

\usepackage{amsmath,amssymb}
\usepackage{graphicx}

\begin{document}

\title{Attempt-time Monte Carlo: an alternative for simulation of
  stochastic jump processes with time-dependent transition rates}

\author{Viktor Holubec$^{1}$}
\author{Petr Chvosta$^1$}
\author{Mario Einax$^{2}$}
\author{Philipp Maass$^{2}$}
\affiliation{
$^{1}$Department of Macromolecular Physics, Faculty of
  Mathematics and Physics, Charles University,
  CZ-18000 Praha, Czech Republic\\  
$^{2}$Fachbereich Physik, Universit\"at Osnabr\"uck,
      Barbarastra\ss e 7, 49076 Osnabr\"uck, Germany}

\date{\today}

\begin{abstract}
  We present a new method for simulating Markovian jump
  processes with time-dependent transitions rates, which avoids the
  transformation of random numbers by inverting time integrals over
  the rates. It relies on constructing a sequence of random time
  points from a homogeneous Poisson process, where the system under
  investigation attempts to change its state with certain
  probabilities. With respect to the underlying master equation the
  method corresponds to an exact formal solution in terms of a Dyson
  series. Different algorithms can be derived from the method and
  their power is demonstrated for a set of interacting two-level
  systems that are periodically driven by an external field. 
\end{abstract}  


\maketitle

\section{Introduction}
\label{sec:1}
Stochastic jump processes with time-dependent transition rates are of
general importance for many applications in physics and chemistry, in
particular for describing the kinetics of chemical reactions
\cite{Gibson/Bruck:2000,Anderson:2007,Astumian:2007} and the
non-equilibrium dynamics of driven systems in statistical mechanics
\cite{Crooks:1999,Seifert:2005,Esposito/VandenBroeck:2010}. With
respect to applications in interdisciplinary fields they play an
important role in connection with queuing theories.

In general a system with $N$ states is considered that at random time
instants performs transitions from one state to another. In case of a
Markovian jump dynamics the probability for the system to change its
state in the time interval $[t,t+\Delta t[$ is independent of the
history and given by $w_{ij}(t)\Delta t+ o(\Delta t)$, where $j$ and
$i\ne j$ are the initial and target state, respectively, and
$w_{ij}(t)$ the corresponding transition rate at time $t$
($w_{jj}(t)=0$). This implies that, if the systems is in the state $j$
at time $t_0$, it will stay in this state until a time $t>t_0$ with
probability $\phi_j(t,t_0)=\exp[-\int_{t_0}^t d\tau\,w_j^{\rm
  tot}(\tau)]$, where $w_j^{\rm tot}(\tau)=\sum_i w_{ij}(\tau)$ is the
total escape rate from state $j$ at time $\tau$. The probability to
perform a transition to the target state $i$ in the time interval
$[t,t+dt[$ then is $w_{ij}(t)\phi_j(t,t_0)dt$, i.\ e.\
\begin{equation}
\psi_{ij}(t,t_0)=
w_{ij}(t)\exp\left[-\int_{t_0}^td\tau\,w_j^{\rm tot}(\tau)\right]
\label{eq:psi}
\end{equation}
is the probability density for the first transition to state $i$ to
occur at time $t$ after the system was in state $j$ at time $t_0$. Any
algorithm that evolves the system according to Eq.~(\ref{eq:psi})
generates stochastic trajectories with the correct path probabilities.

The first algorithm of this kind was developed by Gillespie
\cite{Gillespie:1978} in generalization of the continuous-time
Monte-Carlo algorithm introduced by Bortz \textit{et al.}
\cite{Bortz/etal:1975}  for time-independent rates.
We call it the reaction time algorithm (RTA) in the following. The
RTA consists of drawing a random time $t$ from the first transition
time probability density
$\psi_j^{\rm tot}(t,t_0)=\sum_i \psi_{ij}(t,t_0) =w_j^{\rm tot}\phi_j(t,t_0)=-\partial_t \phi_j(t,t_0)$ to any
other state $i\ne j$, and a subsequent random selection of the target
state $i$ with probability $w_{ij}(t)/w_j^{\rm tot}(t)$.
In practice these two
steps can be performed by generating two uncorrelated
and uniformly distributed random numbers $r_1$, $r_2$ in the
unit interval $[0,1[$ with some random number generator,
where the first is used to specify the transition time $t$ via
\begin{equation}
W_j(t,t_0)=\int_{t_0}^t d\tau w_j^{\rm tot}(\tau)=-\log(1-r_1)
\label{eq:t-selection}
\end{equation}
and the second is used to select the target state $i$ by requiring
\begin{equation}
\sum_{k=1}^{i-1}\frac{w_{kj}(t)}{w_j^{\rm tot}(t)}
\le r_2 < \sum_{k=1}^i\frac{w_{kj}(t)}{w_j^{\rm tot}(t)}
\label{eq:target-selection}
\end{equation}
Both steps, however, lead to some unpleasant problems in the
practical realization.

The first step according to Eq.~(\ref{eq:t-selection}) requires the
calculation of $W_j(t,t_0)$ and the determination of its inverse
$\tilde W_j(.,t_0)$ with respect to $t$ in order to obtain the
transition time $t=\tilde W_j(-\log(1-r_1),t_0)$. While this is always
possible, since $w_j^{\rm tot}>0$ and accordingly $W_j(t,t_0)$ is a
monotonously increasing function of $t$, it can be CPU time consuming
in case $W_j(t,t_0)$ cannot be explicitly given in an analytical form
and one needs to implement a root finding procedure.

The second step according to Eq.~(\ref{eq:target-selection}) can be
cumbersome in case there are many states ($N$ large) and a systematic
grouping of the $w_{ij}(t)$ to only a few classes is not possible.
This situation in particular applies to many-particle systems, where
$N$ typically grows exponentially with the number of particles, and
the interactions (or a coupling to spatially inhomogeneous
time-dependent external fields) can lead to a large number of
different transitions rates. Moreover, even for systems with simple
interactions (as, for example, Ising spin systems), where a grouping
is in principle possible, the subdivision of the unit interval
underlying Eq.~(\ref{eq:target-selection}) cannot be strongly
simplified for time-dependent rates.

A way to circumvent Eq.~(\ref{eq:target-selection}) is the use of the
First Reaction Time Algorithm (FRTA) for time dependent rates
\cite{Jansen:1995}, or modifications of it \cite{Anderson:2007}. In the
FRTA one draws random first transition times $t_k$ from the
probability densities $\psi_{kj}(t_k,t_0)
=w_{kj}(t_k)\exp[-\int_{t_0}^{t_k} d\tau\,w_{kj}(\tau)]$ for the
individual transitions to each of the target states $k$ and performs
the transition $i$ with the smallest $t_i=\min_k\{t_k\}$ at time
$t_i$. This is statistically equivalent to the RTA, since for the
given initial state $j$, the possible transitions to all target states
are independent of each other. In short-range interacting systems, in
particular, many of the random times $t_k$ can be kept for determining
the next transition following $i$. In fact, all transitions from the
new state $i$ to target states $k$ can be kept for which
$w_{ki}(\tau)=w_{kj}(\tau)$ for $\tau>t$ (see
Ref.~\cite{Einax/Maass:2009} for details). However, the random times
$t_k$ need to be drawn from $\psi_{kj}(t_k,t_0)$ and this
unfortunately involves the same problems as discussed above in
connection with Eq.~(\ref{eq:t-selection}).

\section{Algorithms}
\label{sec:2}
We now present a new ``attempt time algorithm'' (ATA) that allows one
to avoid the problems associated with the generation of the transition
time in Eq.~(\ref{eq:t-selection}). Starting with the system in state
$j$ at time $t_0$ as before, one first considers a large time interval
$T$ and determines a number $\mu_j^{\rm tot}$ satisfying
\begin{equation}
\mu_j^{\rm tot}\ge \max_{t_0\le\tau\le t_0+T}\{w_j^{\rm tot}(\tau)\}\,\,.
\label{eq:mutot}
\end{equation}
In general this can by done easily, since $w_j^{\rm tot}(\tau)$ is a
known function. In particular for bounded transition rates it poses no
difficulty, as, for example, in the case of Glauber rates or a
periodic external driving, where $T$ could be chosen as the time
period. If an unlimited growth of $w_j^{\rm tot}$ with time were
present (an unphysical situation for long times), $T$ can be chosen
self-consistently by requiring that the time $t$ for the next
transition to another state $i\ne j$ (see below) must be smaller than
$t_0+T$.

Next an attempt time interval $\Delta t_1$ is drawn from the
exponential density $F_j(\Delta t_1)=\mu_j^{\rm tot} \exp(-\mu_j^{\rm
  tot}\Delta t_1)$ and the resulting attempt transition time
$t_1=t_0+\Delta t_1$ is rejected with probability $p_j^{\rm
  rej}(t_1)=1-w_j^{\rm tot}(t_1)/\mu_j^{\rm tot}$. If it is rejected,
a further attempt time interval $\Delta t_2$ is drawn from $F_j(\Delta
t_2)$, corresponding to an attempt transition time $t_2=t_1+\Delta
t_2$, and so on until an attempt time $t<t_0+T$ is eventually
accepted. Then a transition to a target state $i$ is performed at time
$t$ with probability $w_{ij}(t)/w_j^{\rm tot}(t)$, using the target
state selection of Eq.~(\ref{eq:target-selection}).

In order to show that this method yields the correct first transition
probability density $\psi_{ij}(t,t_0)$ from Eq.~(\ref{eq:psi}), let us
first consider a sequence, where exactly $n\ge0$ attempts at some
times $t_1<\ldots<t_n$ are rejected and then the $(n+1)$th
attempt leads to a transition to the target state $i$ in the time
interval $[t,t+dt[$. The corresponding probability density
$\psi_{ij}^{(n)}(t,t_0)$ is given by
\begin{align}
\label{eq:ata-proof-1}
\psi_{ij}^{(n)}&(t,t_0)=
\int_{t_0}^t dt_n\int_{t_0}^{t_{n-1}} dt_{n-1}\ldots\int_{t_0}^{t_{2}} dt_1
\frac{w_{ij}(t)}{w_j^{\rm tot}(t)} \nonumber \\
&\hspace*{-2em} \times \left[1-p_j^{\rm rej}(t)\right]F_j(t-t_n)
\prod_{m=1}^n p_j^{\rm rej}(t_m)F_j(t_m-t_{m-1}) \\
&=\frac{w_{ij}(t)e^{-\mu_j^{\rm tot}(t-t_0)}}{n!}
\left[\int_{t_0}^td\tau\,\mu_j^{\rm tot}p_j^{{\rm rej}}(\tau)\right]^n\nonumber\\
&=\frac{w_{ij}(t)e^{-\mu_j^{\rm tot}(t-t_0)}}{n!}
\left[\mu_j^{\rm tot}(t-t_0)-\int_{t_0}^td\tau\,w_j^{\rm tot}(\tau)\right]^n.
\nonumber
\end{align}
Summing over all possible $n$ hence yields
\begin{equation}
\psi_j(t,t_0)=\sum_{n=0}^\infty\psi_{ij}^{(n)}(t,t_0)=
w_{ij}(t)\exp\left[-\int_{t_0}^td\tau\,w_j^{\rm tot}(\tau)\right]
\label{eq:ata-proof-2}
\end{equation}
from Eq.~(\ref{eq:psi}).

It is clear that for avoiding the root finding of
Eq.~(\ref{eq:t-selection}) by use of the ATA, one has to pay the price
for introducing rejections. If the typical number of rejections can be
kept small and an explicit analytical expression for $t$ cannot be
derived from Eq.~(\ref{eq:t-selection}), the ATA should become
favorable in comparison to the RTA. Moreover, the ATA can be
implemented in a software routine independent of the special form of
the $w_{ij}(\tau)$ for applicants who are not interested to invest
special thoughts on how to solve Eq.~(\ref{eq:t-selection}).

One may object that the ATA still entails the problem connected with
the cumbersome target state selection by
Eq.~(\ref{eq:target-selection}). However, as the RTA has the first
reaction variant FRTA, the ATA has a first attempt variant. In this
first attempt time algorithm (FATA) one first determines, instead of
$\mu_j^{\rm tot}$ from Eq.~(\ref{eq:mutot}), upper bounds for the
individual transitions to all target states $k\ne j$ ($\mu_{jj}=0$),
\begin{equation}
\mu_{kj}\ge\max_{t_0\le\tau\le t_0+T}\{w_{kj}(\tau)\}\,\,.
\label{eq:mukj}
\end{equation}
Thereon random time intervals $\Delta t_k$ are drawn from
$F_{kj}(\Delta t_k)=\mu_{kj}\exp(-\mu_{kj}\Delta t_k)$, yielding
corresponding attempt transition times $t_k^{(1)}=t_0+\Delta t_k$. The
transition to the target state $k'$ with the minimal
$t_{k'}^{(1)}=\min_k\{t_k^{(1)}\}=t_1$ is attempted and rejected with
probability
$p_{k'k}^{\rm rej}(t_{k'}^{(1)})=1-w_{k'k}(t_{k'}^{(1)})/\mu_{k'k}$. If it is
rejected, a further time interval $\Delta t_{k'}^{(2)}$ is drawn from
$F_{k'j}(\Delta t_{k'}^{(2)})$, yielding
$t_{k'}^{(2)}=t_{k'}^{(1)}+\Delta t_{k'}^{(2)}$, while the other
attempt transition times are kept, $t_k^{(2)}=t_k^{(1)}$ for $k\ne k'$
(it is not necessary to draw new time intervals for these target
states due to the absence of memory in the Poisson process). The
target state $k''$ with the new minimal
$t_{k''}^{(2)}=\min_k\{t_k^{(2)}\}=t_2$ is then attempted and so on
until eventually a transition to a target state $i$ is accepted at a
time $t<t_0+T$. The determination of the minimal times can be done
effectively by keeping an ordered stack of the attempt times.
Furthermore, as in the FRTA, one can, after a successful transition to
a target state $i$ at time $t$, keep the (last updated) attempt times
$t_k$ for all target states that are not affected by this transition
(i.\ e.\ for which $w_{ki}(\tau)=w_{kj}(\tau)$ for $\tau\ge t$).
Overall one can view the procedure implied by the FATA as that each
state $k$ has a next attempt time $t_k$ (with $t_j=\infty$ if the
system is in state $j$) and that the next attempt is made to the
target state with the minimal $t_k$. After each attempt, updates of
some of the $t_k$ are made as described above in dependence of whether
the attempt was rejected or accepted.

In order to prove that the FATA gives the $\psi_j(t,t_0)$ from
Eq.~(\ref{eq:psi}), we show that the probability densities
$\chi_{ij}(t,t_n)=[w_{ij}(t)/w_j^{\rm tot}(t)](1-p_j^{\rm
  rej}(t))F_j(t-t_n)=w_{ij}(t)\exp[-\mu_j^{\rm tot}(t-t_n)]$ and
$\eta_j(t_m,t_{m-1})= p_j^{\rm rej}(t_m)F_j(t_m-t_{m-1})=[\mu_j^{\rm
  tot}-w_j^{\rm tot}] \exp[-\mu_j^{\rm tot}(t_m-t_{m-1})]$ appearing
in Eq.~(\ref{eq:ata-proof-1}) are generated, if we set $\mu_j^{\rm
  tot}=\sum_k \mu_{kj}$ (note that Eq.~(\ref{eq:mutot}) is
automatically satisfied by this choice). These probability densities
have the following meaning: $\chi_{ij}(t,t_n)dt$ is the probability
that, if the system is in state $j$ at time $t_n$, the next attempt to
a target state occurs in the time interval $[t,t+dt[$, the attempt is accepted,
and it changes the state from $j$ to $i$; $\eta_j(t_m,t_{m-1})dt_m$ is
the probability that, after the attempt time $t_m$, the next attempt
occurs in $[t_m,t_m+dt_m[$ with $t_m>t_{m-1}$ and is rejected.

In the FATA the probability $\kappa_{lj}(t_m,t_{m-1})dt_m$ that, when
starting at time $t_{m-1}$, the next attempt is occurring in
$[t_m,t_m+dt_m[$ to a target state $l$ is given by
\begin{align}
\kappa_{lj}(t_m,t_{m-1})&=
\mu_{lj}\exp[-\mu_{lj}(t_m-t_{m-1})]\nonumber \\
& \hspace*{0.3cm} \times \prod_{k\ne l} \int_{t_m-t_{m-1}}^\infty
d\tau\,\mu_{kj}\exp(-\mu_{kj}\tau)\nonumber\\
&=\mu_{lj}\exp[-\mu_j^{\rm tot}(t_m-t_{m-1})]\,\,.
\label{eq:kappa}
\end{align}
The product ensures that $t_m$ is the minimal time (the lower bound in
the integral can be set equal to $(t_m-t_{m-1})$ for all $k\ne l$ due
to the absence of memory in the Poisson process). The probability that
this attempted transition is rejected is $p_{lj}^{\rm
  rej}(t_m)=1-w_{lj}(t_m)/\mu_{lj}$ and accordingly, by summing over
all target states $l$, we obtain
\begin{align}
\eta_j(t_m,t_{m-1})&=\sum_l p_{lj}^{\rm
rej}(t_m)\mu_{lj}\exp[-\mu_j^{\rm tot})(t_m-t_{m-1})]\nonumber\\
&=[\mu_j^{\rm tot}-w_j^{\rm tot}(t_m)]\exp[-\mu_j^{\rm tot}(t_m-t_{m-1})]
\label{eq:eta}
\end{align}
in agreement with the expression appearing in
Eq.~(\ref{eq:ata-proof-1}). Furthermore, when starting from time
$t_n$, the probability density $\chi_{ij}(t,t_n)$ referring to the
joint probability that the next attempted transition occurs in
$[t,t+dt[$ to state $i$ and is accepted is given by
\begin{equation}
\chi_{ij}(t,t_n)=\frac{w_ {ij}(t)}{\mu_{ij}}\kappa_{ij}(t,t_n)=
w_{ij}(t)\exp[-\mu_j^{\rm tot}(t-t_{n})]\,\,.
\label{eq:chi}
\end{equation}
Hence one recovers the decomposition in Eq.~(\ref{eq:ata-proof-1})
with $\mu_j^{\rm tot}=\sum_k \mu_{kj}$.

Before discussing an example, it is instructive to see how the ATA
(and RTA) can be associated with a solution of the underlying master
equation
\begin{equation}
  \frac{\partial}{\partial t}\,\mathbb{G}(t,t')=-\mathbb{M}(t)\,
  \mathbb{G}(t,t')\,\,,\qquad\mathbb{G}(t',t')=\mathbb{I}
\label{eq:master}
\end{equation}
where $\mathbb{G}(t,t')$ is the matrix of transition probabilities
$G_{ij}(t,t')$ for the system to be in state $i$ at time $t$ if it was
in state $j$ at time $t'\le t$, and $\mathbb{M}(t)$ is the transition
rate matrix with elements $M_{ij}(t)=-w_{ij}(t)$ for $i\ne j$ and
$M_{jj}(t)=-\sum_{i\ne j} M_{ij}(t)=w_j^{\rm tot}(t)$. Let us
decompose $\mathbb{M}(t)$ as $\mathbb{M}(t)=\mathbb{D}+\mathbb{A}(t)$,
where $\mathbb{D}=\mathrm{diag}\left\{ \mu_1^{\rm
    tot},\ldots,\mu_N^{\rm tot}\right\}$. If $\mathbb{A}(t)$ were
missing, the solution of the master equation (\ref{eq:master}) would
be $\mathbb{G}_0(t,t')= \mathrm{diag}\left\{\exp(-\mu_1^{\rm
    tot}(t-t'), \ldots,\exp(-\mu_N^{\rm tot}(t-t')\right\}$. Hence,
when introducing
$\tilde{\mathbb{A}}(t,t')=\mathbb{G}_0^{-1}(t,t')\mathbb{A}(t)
\mathbb{G}_0(t,t')=\mathbb{G}_0(t',t)\mathbb{A}(t) \mathbb{G}_0(t,t')$
in the ``interaction picture'', the solution of the master equation
can be written as
\begin{align}
  \mathbb{G}(t,t')&=\mathbb{G}_0(t,t')\left[
    \mathbb{I} +\int_{t'}^t dt_1 \tilde{\mathbb{A}}(t_1,t') \right. \nonumber \\
  & \hspace*{0.3cm} + \left.\int_{t'}^t dt_2 \int_{t'}^{t_2} dt_1
    \tilde{\mathbb{A}}(t_2,t')
    \tilde{\mathbb{A}}(t_1,t')+\ldots\right]
\label{eq:master-solution-1}
\end{align}
Inserting $\mathbb{I}=\mathbb{D}^{-1}\mathbb{D}$ after each matrix
$\tilde{\mathbb{A}}$, one arrives at
\begin{align}
  &\mathbb{G}(t,t')=\mathbb{G}_0(t,t')
  +\int_{t'}^t dt_1 \mathbb{G}_0(t,t_1)\mathbb{B}(t_1)F_0(t_1,t') \nonumber \\
  &{}+\int_{t'}^t dt_2 \int_{t'}^{t_2} dt_1
  \mathbb{G}_0(t,t_2)\mathbb{B}(t_2)\mathbb{F}_0(t_2,t_1)
  \mathbb{B}(t_1)\mathbb{F}_0(t_1,t') \nonumber \\ &+\ldots
\label{eq:master-solution-2}
\end{align}
where
$\mathbb{F}_0(t,t')=\mathbb{D}\,\mathbb{G}_0(t,t')=
\mathrm{diag}\{\mu_1^{\rm tot}\exp[-\mu_1^{\rm tot}(t-t')]
,\ldots,$
$\mu_N^{\rm tot}\exp[-\mu_N^{\rm tot}(t-t')]\}$, and
$\mathbb{B}(t)=\mathbb{A}(t)\mathbb{D}^{-1}$ has the matrix elements
$B_{ij}(t)=-w_{ij}(t)/\mu_j^{\rm tot}$ for $i\ne j$ and
$B_{jj}(t)=1-w_j^{\rm tot}(t)/\mu_j^{\rm tot}$.

Equation~(\ref{eq:master-solution-2}) resembles the ATA: The
transition probabilities $G_{ij}(t,t')$ are decomposed into paths with
an arbitrary number $n=0,1,2,\ldots$ of ``Poisson points'', where
transitions are attempted. The times between successive attempted
transitions are exponentially distributed according to the matrix
elements of $\mathbb{F}_0$ and the attempted transitions are accepted
or rejected according to the probabilities encoded in the diagonal and
non-diagonal elements of the $\mathbb{B}$ matrix, respectively. The
$\mathbb{G}_0$ entering Eq.~(\ref{eq:master-solution-2}) takes
care that after the last attempt in a path with exactly $n$ attempted
transitions no further attempt occurs and the system remains in the
target state $i$. The RTA can be associated with an analogous formal
solution of the master equation if one replaces $\mathbb{G}_0(t,t')$
by $\mathbb{G}_0^{\rm RTA}(t,t')=
\mathrm{diag}\left\{ w_1^{\rm tot}(t)\exp[-\int_{t'}^td\tau\,w_1^{\rm tot}(\tau)]
,\ldots,\right.$
$\left.
w_N^{\rm tot}(t)\exp[-\int_{t'}^td\tau\,w_N^{\rm tot}(\tau)]\right\}$
and
$\mathbb{B}(t)$ by $\mathbb{B^{\rm RTA}}(t)$ with elements
$B_{ij}^{\rm RTA}(t)=(1-\delta_{ij})w_{ij}(t)/w_j^{\rm tot}(t)$ (the
diagonal elements are zero since the RTA is rejection-free).

\section{Example}
\label{sec:example}
Let us now demonstrate the implementation of the FATA in an example. To
this end we consider three mutually coupled two-level systems that
are periodically driven. For an arbitrary given $i$, $i=1,2,3$, the state $|\,i,\pm\,\rangle$
has the energy $\pm E(t)$.
The occupancy of the state $|\,i,\pm\,\rangle$ is specified by the occupation number $n_i=\pm 1$.
For example, if $n_i=-1$, the $i$-th two level system resides in the state $|\,i,-\,\rangle$ and
it possesses the energy $-E(t)$.
The coupling is described by the (positive) interaction parameter $V$.
The total energy of the three coupled two-level systems is given by the expression
\begin{equation}
H(\mathbf{n},t)=V(n_1n_2+n_1n_3+n_2n_3)+E(t)\sum_{i=1}^{3}n_{i}\,\,,
\label{eq:h}
\end{equation}
where $\mathbf{n}=(n_{1},n_{2},n_{3})$ specifies the microstate of the compound system.
The periodic driving is considered to change energies of the individual two-level
systems as
\begin{equation}
E(t)=\frac{\Delta E}{2}\,\sin(\omega t)\,,
\label{eq:e}
\end{equation}
where $\Delta E>0$ is the amplitude of modulation and $\omega$ its
frequency. Due to contact of the compound system with a heat
reservoir at temperature $T$, transitions between its microstates occur.
Assume that in the initial state $\mathbf{m}$ one and only one occupation number differs from the corresponding occupation number
in the final state $\mathbf{n}$. Then instantaneous value of the detailed-balanced Glauber jump rates
connecting these two states reads
\begin{align}
w(\mathbf{m}\rightarrow\mathbf{n},t)=
\frac{\nu}{1+\exp\left\{\beta\left[H(\mathbf{n},t)-H(\mathbf{m},t)\right]\right\}}\,\,.
\label{eq:w}
\end{align}
The other pairs of microstates are not connected, that is, the transition rates between them vanish.
In the above expression, $\nu$ designates an attempt frequency, and $\beta$ is the inverse temperature.
In the following we will use $k_{\rm B}T$ as our energy unit and $\nu^{-1}$ as our time unit.
\begin{figure}[t!] 
\begin{center}
\includegraphics[width=0.45\textwidth]{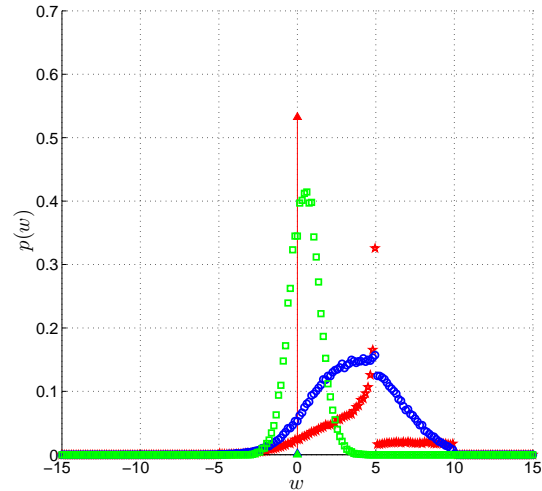}
\caption{Work distributions $p(w)$ as obtained from the FATA for $\Delta E=V=5$ and $\omega=0.1$ (squares, green color),
1 (circles, blue color), and $\omega=10$ (stars, red color).}
\label{fig:fig1}
\end{center}
\end{figure}

In current research of non-equilibrium systems, in particular of
processes in small molecular systems, the investigation of
distributions of microscopic work receives much attention. Among
others, this is largely motivated by questions concerning the
optimization of processes, and by the connection of the work
distributions to fluctuation theorems. These theorems allow one to
obtain equilibrium thermodynamic quantities from the study of
non-equilibrium processes and they are useful for getting a deeper
insight into the manifestation of the second law of thermodynamics.
At the same time, the analytical expressions for the work distribution are
rarely attainable (one exception is reported in \cite{Chvosta:2007}).
It is therefore interesting to see how the FATA can be employed for
studies in this research field. To be specific, we focus on the
stationary state and calculate work distributions within one period of
the external driving. For these distributions we check the detailed
fluctuation theorem of Crooks \cite{Crooks:1999}, as generalized by
Hatano and Sasa \cite{Hatano/Sasa:2001} to steady states (for a nice
summary of different forms of detailed and integral fluctuation
theorems, see \cite{Esposito/VandenBroeck:2010}).

In our model, due to the possibility of thermally activated transitions between the eight microstates, the
state vector $\mathbf{n}$ must be understood as a stochastic process. We designate it as $\mathbf{n}(t)$, and let
$\mathbf{n}^{{\rm tr}}(t)$ denotes its arbitrary fixed realization. The instantaneous energy of the compound system along this
realization is then $H(\mathbf{n}^{{\rm tr}}(t),t)$. The work done on the system during the $m$th period $[m\tau,(m+1)\tau]$, $\tau=2\pi/\omega$,
if the system evolves along the realization in question, is given by
\begin{align}
w^{{\rm tr}}_{m}&=\int_{m\tau}^{(m+1)\tau}dt\,\frac{\partial}{\partial t}H(\mathbf{n}^{{\rm tr}}(t),t)
\nonumber\\
&=
\frac{\omega\Delta E}{2}\sum_{i=1}^3\int_{m\tau}^{(m+1)\tau}dt\, n^{{\rm tr}}_{i}(t)\cos(\omega t)\,\,.
\label{eq:work}
\end{align}
In the stationary limit $m\to\infty$ ($m\gg1$) we can drop the
index $m$. According to the detailed fluctuation theorem, the work
distribution $p(w)$ should, in our case
(time-symmetric situation with respect to the initial microstate distribution for starting forward and backward paths),
obey the relation $p(w)\exp(-w)=p(-w)$.

Figure~\ref{fig:fig1} shows the results for $p(w)$ obtained from the
FATA for $\Delta E=V=5$, and three different frequencies $\omega=0.1$,
1, and 10.
First, we let the system evolve during the $N_{{\rm ini}}=1$ ($\omega=0.1$),
$N_{{\rm ini}}=3$ ($\omega=1$), $N_{{\rm ini}}=9$ ($\omega=10$) periods
to reach the stationary state.
Subsequently, the work values $w^{{\rm tr}}$ according to Eq.~(\ref{eq:work})
were sampled over $N=10^{4}$ ($\omega=0.1$), $N=10^{5}$
($\omega=1$), and $N=10^{5}$ ($\omega=10$) periods.

With decreasing
$\omega$, the maxima of the work distributions in Fig.~\ref{fig:fig1}
shift toward $w=0$, and $\delta$-singularities, marked by the vertical
lines, receive less weight. These $\delta-$singularities are
associated with stochastic trajectories of the system, where no
transitions occur within a period of the driving. For $\omega=0.1$,
$p(w)$ is already close to the Gaussian fluctuation regime.

\begin{figure}[t!] 
\begin{center}
\includegraphics[width=0.45\textwidth]{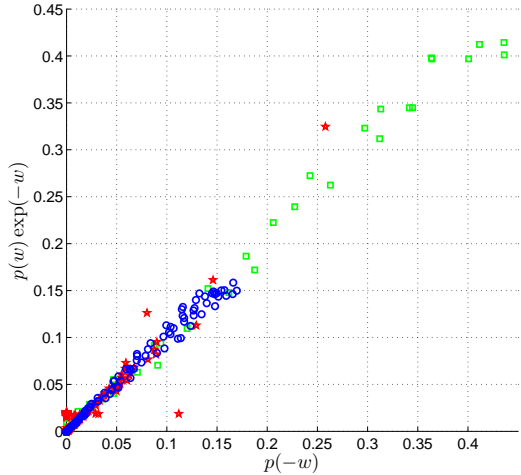}
\caption{Check of the detailed fluctuation theorem $p(w)\exp(-w)=p(-w)$
  for the work distributions shown in Fig.~\ref{fig:fig1}. The same
  symbols/colors are used for the three different frequencies as in
  Fig.~\ref{fig:fig1}.}
\label{fig:fig2}
\end{center}
\end{figure}

In Fig.~\ref{fig:fig2} we show that the work distributions from
Fig.~\ref{fig:fig1} indeed fulfill the detailed fluctuation theorem.
This demonstrates that the FATA successfully generates system
trajectories with the correct statistics of the stochastic process.

\section{Summary}
\label{sec:summary}
In summary, we have presented new simulation algorithms for Markovian
jump processes with time-dependent transition rates, which avoid the
often cumbersome or unhandy calculation of inverse functions. The ATA
and FATA rely on the construction of a series of Poisson points, where
transitions are attempted and rejected with certain probabilities. As
a consequence, both algorithms are easy to implement, and their
efficiency will be good as long as the number of rejections can be
kept small. For complex interacting systems, the FATA has the same
merits as the FRTA with respect to the FRA. Both the ATA and FATA
generate exact realizations of the stochastic process. Their
connection to perturbative solutions of the underlying master equation
may allow one to include in future work also non-Markovian features of
a stochastic dynamics by letting the rejection
probabilities to depend
on the history \cite{Chvosta:1999}. Compared to the RTA and FRTA, the new algorithms
should in particular be favorable, when considering periodically
driven systems with interactions. Such systems are of much current
interest in the study of non-equilibrium stationary states and we thus
hope that our findings will help to investigate them more conveniently
and efficiently.

\begin{acknowledgements}
Support of this work by the Ministry of Education of the Czech Republic (project No. MSM 0021620835), by the Grant Agency of the Charles University (grant No. 143610) and by the project
SVV - 2010 - 261 301 of the Charles University in Prague is gratefully acknowledged.
\end{acknowledgements}


\end{document}